\documentclass{JINST}
\RequirePackage{xspace}
\def\epem {\ensuremath{e^+e^-}\xspace}
\def\gg {\ensuremath{\gamma \gamma}\xspace}
\def\ge {\ensuremath{\gamma e}\xspace}

\def\invfb {\ensuremath{\mathrm{fb}^{-1}}\xspace}

\def\t-t {\ensuremath{{\rm   tt}}\xspace}

\newcommand{\mev}{\ensuremath{\mathrm{\,Me\kern -0.1em V}}\xspace}
\newcommand{\mevcc}{\ensuremath{{\mathrm{\,Me\kern -0.1em V\!/}c^2}}\xspace}
\newcommand{\gev}{\ensuremath{\mathrm{\,Ge\kern -0.1em V}}\xspace}
\newcommand{\gevcc}{\ensuremath{{\mathrm{\,Ge\kern -0.1em V\!/}c^2}}\xspace}
\newcommand{\tev}{\ensuremath{\mathrm{\,Te\kern -0.1em V}}\xspace}
\newcommand{\tevcc}{\ensuremath{{\mathrm{\,Te\kern -0.1em V\!/}c^2}}\xspace}
\newcommand{\ev}{\ensuremath{\mathrm{\,e\kern -0.1em V}}\xspace}

\def\mum  {\ensuremath{{\,\mu\rm m}}\xspace}

\def\cms  {\ensuremath{{\rm \,cm}^{-2} {\rm s}^{-1}}\xspace}

\newcommand{\EPEM}{\ensuremath{e^+e^{-}}\xspace}

\newcommand{\GG}{\ensuremath{\gamma\gamma}\xspace}

\newcommand{\GEV}{\ensuremath{GeV}\xspace}

\newcommand{\MKM}{\ensuremath{\mu m}\xspace}

\newcommand{\be}{\begin{equation}}
\newcommand{\ee}{\end{equation}}
\newcommand{\bc}{\begin{center}}
\newcommand{\ec}{\end{center}}
\newcommand{\bi}{\begin{itemize}}
\newcommand{\ei}{\end{itemize}}
\newcommand{\ben}{\begin{enumerate}}
\newcommand{\een}{\end{enumerate}}

\title{Photon collider Higgs factories }

\author{V.~I.~Telnov  \\
Budker Institute of Nuclear Physics,\\
Novosibirsk State University, \\ 630090, Novosibirsk, Russia\\
E-mail: \email{telnov@inp.nsk.su}}

\abstract{The discovery of the Higgs boson (and still nothing else)
have triggered appearance of many proposals of Higgs
factories for precision measurement of the Higgs
properties. Among them there are several projects of photon colliders (PC)
without \epem  in addition to PLC based on \epem linear colliders ILC and CLIC.
In this paper, following a brief discussion of Higgs factories physics program  I give an overview of photon colliders based on linear colliders ILC and CLIC,
and of the recently proposed photon-collider Higgs factories with no \EPEM\ collision option based on recirculation linacs in ring tunnels.}


\begin{document}

\section{Introduction} In the middle of 2012, two detectors at the LHC announced the discovery of a new particle with the mass of about 126 \gevcc, with properties consistent with those predicted for the Standard Model Higgs boson. This discovery, without a doubt, is a huge success. Physicists around the world had been waiting for many years for the first round of LHC discoveries in order to decide what the next HEP projects should be.

 Since 1990-th
 the HEP community was unanimous that the next large HEP project should be a linear collider (LC) with the energy $2E_0$=500--1000 \gev. At present, two linear collider projects remain: ILC ($2E_0$=250--1000 \gev)~\cite{ILC} and CLIC ($2E_0$=350--3000 \gev)~\cite{CLIC}. The ILC team has already prepared a Technical Design Report and almost ready to start construction (in Japan), while the CLIC team issued a Conceptual Design, with a Technical Design to be ready a few years later. Due to high costs, it is clear that no more than one LC can be build---but which one (or no one)?

Up to now, the LHC has found only the Higgs boson---and nothing else below approximately one \tevcc: no supersymmetry, no dark matter particles, not a hint of anything else. It is not excluded that new physics will yet be found at LHC as higher statistics are accumulated and as LHC ramps up to its full design energy of 14 TeV.  This means that the LC decision could be made no earlier than 2018. The physics motivation for an energy-frontier LC is no longer as strong as before because we know that the energy region below 1 TeV is not nearly as rich as had been expected. Are there any other strategies HEP could follow? For example,  building a low-energy facility for the detailed study of the Higgs boson while leaving the energy frontier to the LHC, the high-luminosity HL-LHC, and to some future, even more high-energy $pp$ or muon collider.

At first sight, even in this scenario, the ILC is the top candidate for a Higgs factory; if not the ILC---what else? At the end on 2011,  A.~Blondel and F.~Zimmermann published an e-print~\cite{Zim} with the proposal of an \epem ring collider in the LHC tunnel, dubbed LEP3, to study the Higgs boson. That e-print triggered a strong renewed interest in \epem ring colliders, because for such a low-mass Higgs boson one needs a ring with the energy $2E_0 = 240 \gev$, which is only somewhat larger than it was at LEP-2 (209 GeV). Soon thereafter, it became clear that it would be preferable to build a ring collider with a radius several times as large as LEP-2's because: a) for a fixed power, the luminosity is proportional to the ring's radius, b) in the future, one can place in the same tunnel a $\sim 100$ \gev $pp$ collider.
 The luminosity of such a ring collider could be several times larger that at the ILC~\cite{Blo,TLEP}. This clearly sounds like a serious long-term HEP strategy. Another option is a muon-collider Higgs factory~\cite{muon}. The technology is not ready yet, but the development of muon colliders is needed in any case for access to the highest energies. There are also suggestions for a ring-type photon collider Higgs factory (without \epem) based on recirculating linacs), though usually photon colliders are considered as a natural add-on to \epem\ linear colliders. The present list of possible Higgs factories, considered at HF2012~\cite{Blo2} is given below~\cite{TelnovHF}

\begin{itemize}
\item Proton colliders
\begin{itemize}
\item a) LHC, b) HL-LHC, c) HE-LHC, d) SHE-LHC, e) VLHC.
\end{itemize}
\item Linear \epem colliders
\begin{itemize}
\item a) ILC, b) CLIC, c) X-band klystron-based.
\end{itemize}
\item Circular \epem colliders
\begin{itemize}
\item a) LEP3, b) TLEP (Triple-size LEP)~\cite{Blo}, c) SuperTRISTAN-40(80),  d) Fermilab
site-filler, e) CHF-1 and CHF-2 (China), f) VLLC~\cite{Sum}.
\end{itemize}
\item Photon colliders
\begin{itemize}
\item a) ILC-based, b) CLIC-based, c) Recirculating linac-based (SAPPHiRE), d) SLC-type.
\end{itemize}
\item Muon collider
\end{itemize}

Below, we consider shortly the physics motivation of each Higgs factory option and then photon colliders in more detail.

\section{Higgs physics}
\noindent
The Higgs boson has been detected at the LHC in multiple channels: $bb$, $\tau\tau$, $\gg$, $WW$, $ZZ$. The measurement of the Higgs mass gave us
the last unknown parameter of the Standard Model (SM). Now, all SM cross sections and branching ratios can be calculated
and compared with the experiment. Any statistically significant deviation would signal the existence of new physics.
What Higgs properties should be measured and what accuracy is needed?

\begin{figure}[!tbp]
     \begin{center}
     \vspace*{-0.2cm}
\includegraphics[width=8.5cm] {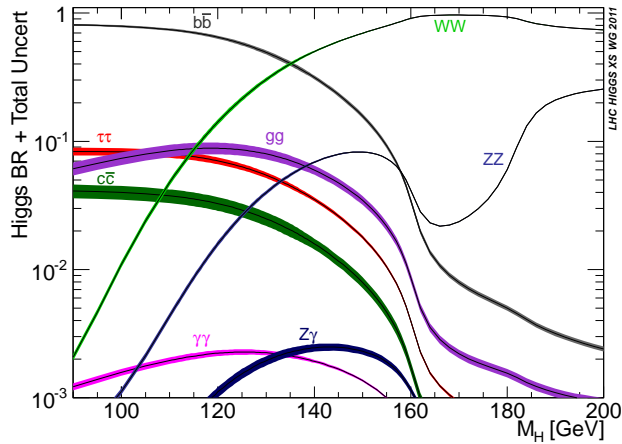}
       \vspace*{-0.7cm}
     \end{center}
     \caption{The Higgs boson branchings. }
   \vspace*{0.0cm}
   \label{h-branch}
   \end{figure}

According to theoretical predictions, new physics appearing at the 1 TeV scale could change the Higgs branching ratios by $\sim 1$--5\%. Detecting such a small discrepancy would require a branching-ratio measurement precision of much better than one percent. The leading branchings of the 126 \gev Higgs boson are shown in Fig.~\ref{h-branch}.  One could also measure the Higgs' branching ratio to $\mu\mu$, which is about 0.022\%; branchings to \epem\ and light quarks are too small to be measured. The total Higgs width is about 4 MeV, it can be measured at \epem and muon colliders (see below). Electron-positron colliders also present a nice possibility to measure the Higgs decay width to invisible states and find  the Higgs total width. The Higgs coupling to the top quark can be measured at both the LHC and high-energy \epem colliders. The Higgs self-coupling is very important but difficult to measure; its measurement requires either the high-luminosity LHC or a high-energy linear \epem collider. Let us consider the Higgs physics accessible at each type of colliders.

\subsection{The LHC as a Higgs factory}
\noindent
The cross sections for Higgs boson production in $pp$ collisions at c.m.s. energies of 7 and 14 TeV can be found elsewhere~\cite{TelnovHF}. The total cross section at 14 TeV is about 57 pb. The expected integrated luminosity is 300 \invfb at the nominal LHC and 3000 \invfb at the high-luminosity HL-LHC, which corresponds to the production of about 20 million and 200 million Higgs bosons, respectively, many more than at any other Higgs factory. The main problems at the LHC are backgrounds and the uncertainties of the initial state and the production mechanisms. The Higgs boson is produced in gluon fusion, vector-boson fusion, and radiation from top quark, $W$ or $Z$ bosons. For each final state, one can identify the initial state by kinematic selection. This way, one can measure the $Htt$ coupling. The Higgs boson can be detected in all final states enumerated above, with the exception of $c\bar{c}$. The total and invisible Higgs widths cannot be measured in $pp$ collisions (or very roughly).

\subsection{\boldmath{Higgs physics at \epem colliders}}
\noindent
 The total cross section at the 240 GeV \epem Higgs factory is about 300 fb (200 fb for unpolarized beams), (see diagrams and cross sections in \cite{TelnovHF}.) Typical luminosity of \epem colliders (LC or rings) is about $10^{34}$ \cms, therefore, the total number of produced Higgs bosons per one year ($10^7$ sec) is about 20000-30000, or about 100000 for life of the experiment. Large circular \epem colliders like TLEP (C$=80$ km) or IHEP (C$=70$ km) can have  luminosities several times greater than at linear colliders. In addition, ring colliders can have several interaction points; therefore, the number of produced Higgs bosons could reach 1 million.

   A unique feature of \epem colliders is the reaction $\epem \to ZH$. By detecting the leptonic decays of the $Z$, one can measure directly all branching ratios and see even invisible Higgs boson decays via the recoil mass.  Measurements of the cross section cross of this process and branching to $ZZ$  gives the total Higgs width. Similar possibilities provides the reaction $\epem \to H\nu\bar{\nu}$.

\subsection{Muon collider Higgs factory}
\noindent
   At muon colliders, the Higgs boson can be produced in the $s$-channel as a single resonance. The cross section of the reaction $\mu^+\mu^- \to H(126)$ is 70 pb. In order to measure directly the Higgs width, the energy spread of muon beams should be comparable to $\Gamma_H\approx 4.2$~MeV. The relative energy spread can be reduced to $3\times 10^{-5}$ by emittance exchange at the expense of transverse emittance. The luminosity will be not high, about $10^{31} \cms$, but due to a high cross section one can produce about 5000 Higgs events per year for one IP. Scanning of the peak gives the Higgs total width with a precision of 3\%; coupling to $\mu\mu$---with accuracy of 1.5\% and the Higgs mass with a 0.1 \mevcc accuracy. All other Higgs parameters can be measured better at \epem colliders thanks to their lower physics and machine backgrounds. Studies are on-going to increase the luminosity at the Higgs up to $10^{32} \cms$.    The muon collider Higgs factory is useful and desirable in all scenarios (with \epem colliders or without) because this technique paves the way to the highest collider energies conceivably achievable by humankind, up to 100 TeV.

\subsection{Higgs physics at photon colliders}

\begin{figure}[tbp]
     \begin{center}
\includegraphics[width=10cm] {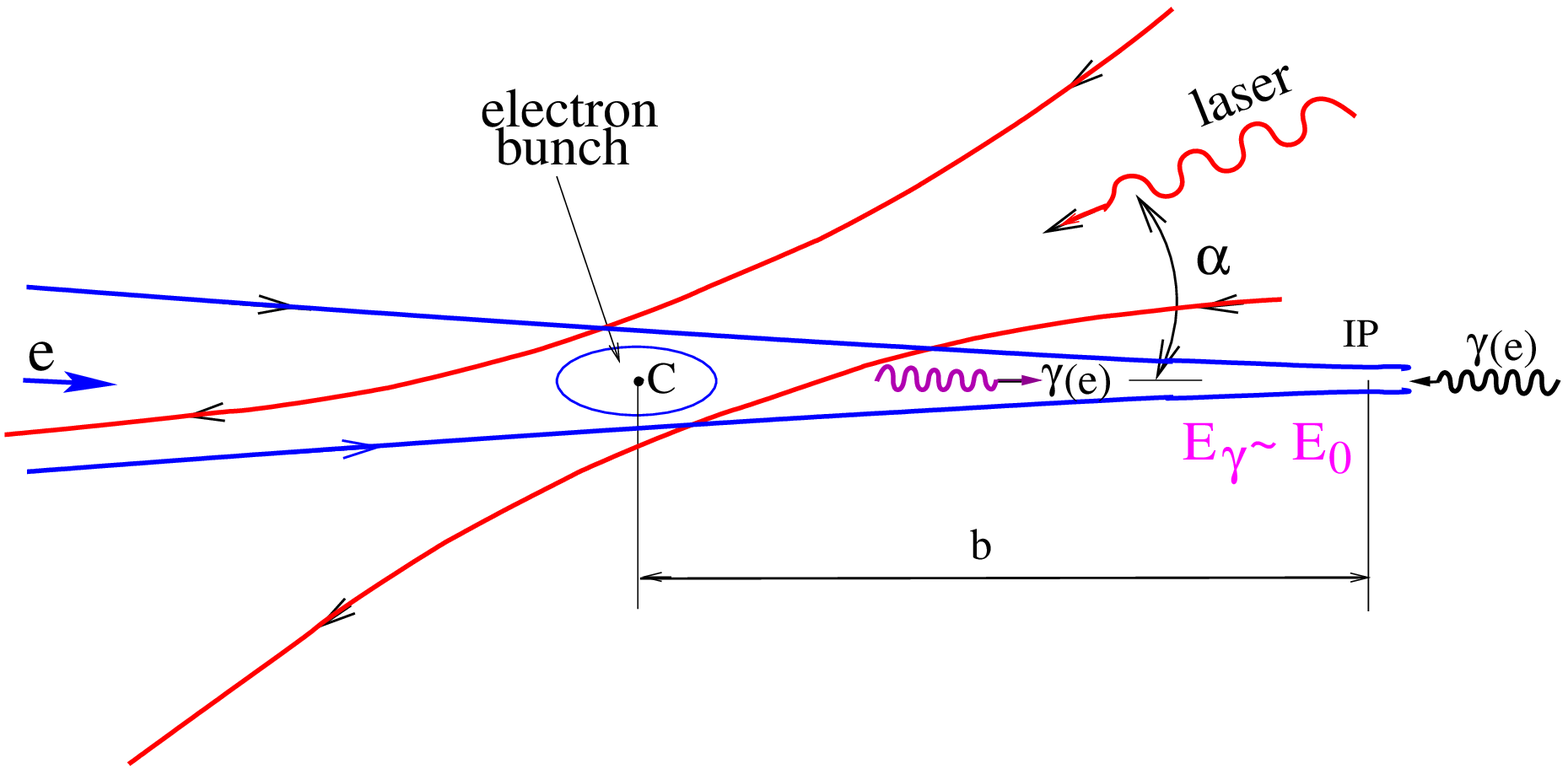} \vspace*{0cm} ~~ \includegraphics[width=5.cm] {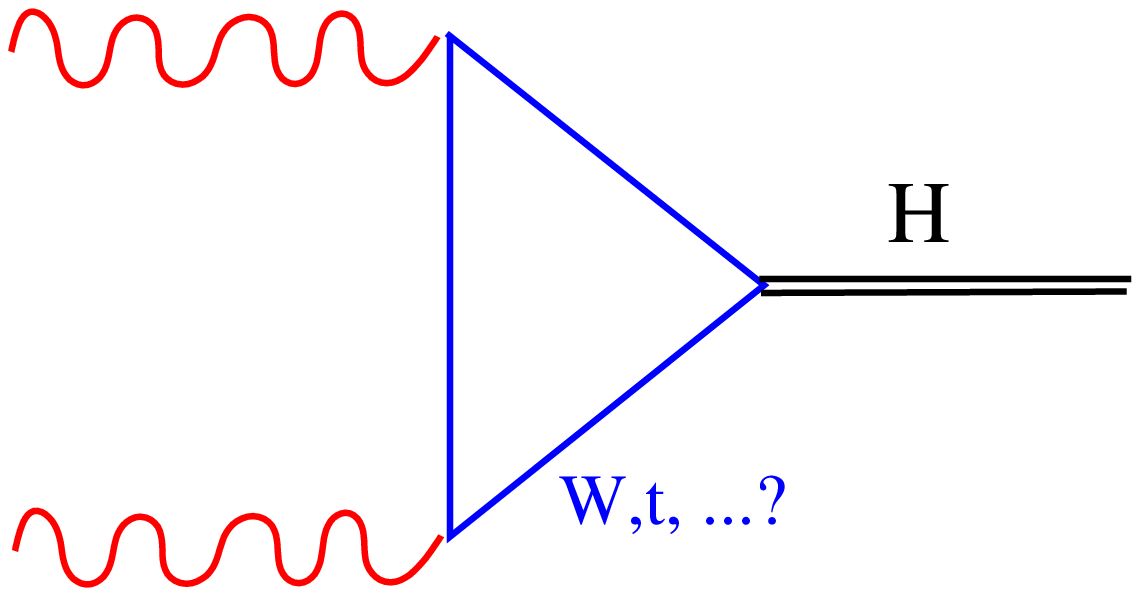}
       \vspace*{0.0cm}
     \end{center}
     \caption{Upper: scheme of \gg, \ge collider ; bottom: the diagram for Higgs production in \gg collsions.}
   \vspace*{0.cm}
   \label{plc}
   \end{figure}

In \gg collisions, the Higgs boson is produced as a single resonance via the loop diagram (Fig.~\ref{plc}) where the leading contributions come from the heaviest charged  particles in the loop: $t$, $W$, $b$.  The measurement of this reaction's cross section can reveal the existence of yet-unknown heavy charged particles that cannot be directly produced at colliders due to their high masses. For monochromatic photons, the cross section would be huge, about 700 pb, a factor of 10 greater than that at a muon collider. Unfortunately, at realistic photon colliders based on Compton backscattering the energy spread of the high-energy peak is about 15\% at half maximum. Even with such an energy spread, the Higgs production rate at the photon collider is comparable to that in \epem collisions (see below).

   The Higgs boson at photon colliders can be observed in the $bb, \gg, WW$ decay modes. The Higgs \gg width $\Gamma_{\gg}$ can be measured with an accuracy of about 2\%, better than at other collider types. However, this requires the knowledge of $\mathrm{BR}(H\to b\bar{b})$, which can be measured with sufficient accuracy only at \epem colliders. Using variable photon polarizations, one can measure the Higgs boson's $CP$ properties. Although the photon collider can produce similar numbers of Higgs bosons as an \epem collider, due to the irreducible QED backgrounds one cannot detect the Higgs in the $cc, \tau\tau, \mu\mu, gg$ modes, measure directly the branchings, and see the invisible decays. Therefore, an \epem collider would be much more powerful for the Higgs study, and the photon collider will be useful only for a number of specific additional measurements---first and foremost, $\Gamma_{\gg}$, which, in fact, could be the most interesting measurement to be done at a photon collider due to its sensitivity to the possible existence of new massive charged particles.




Let us compare the strengths of \EPEM and \GG colliders in the study of the Higgs boson.
The photon collider can measure $\Gamma(H\to \GG)\times Br(H\to bb,ZZ,WW,\GG)$
and, using linearly polarized photons, the Higgs' $CP$ properties.
In order to extract $\Gamma(H \to \GG$), one needs the value of $Br(H \to
bb)$  from an \EPEM\ collider. In \EPEM\ collisions, one can measure $Br(H\to bb,
cc, gg, WW, ZZ, \mu\mu,$ $\mathrm{invisible}$), $\Gamma_{\mathrm{tot}}$. The process $\EPEM \to ZH$ with $Z$ tagging
allows the measurement of the absolute values of branching fractions, including Higgs decays
to $\tau\tau, \mu\mu, cc$, which are not accessible in \GG\ collisions due to a large QED background.

  The rate of Higgs boson production in \GG\ collisons~\cite{TESLATDR}
  \be
  \dot{N}_H=L_{ee} \times \frac{dL_{0,\gg}}{dW_{\gg}L_{ee}}\frac{4\pi^2\Gamma_{\gg}}{M_H^2}(1+\lambda_1\lambda_2+CP*l_1l_2 cos2\varphi)=
  L_{ee}\sigma
  \ee \\[-0.5cm]
  $$    \sigma=\frac{0.98\cdot 10^{-35}}{2E_0[\GEV]} \frac{dL_{0,\gg}}{dz L_{ee}} (1+\lambda_1\lambda_2+CP*l_1l_2 cos2\varphi), \; \mbox{cm}$$
where $L_{ee}$ is the geometric $ee$ luminosity, $L_{0,\gg}$ is the \GG\ luminosity at total helicity zero,
$z=W_{\gg}/2E_0$, $\lambda_{1,2}$ and  $l_{1,2}$ are the helicities and linear polarizations of the high-energy photons,
$\varphi$ is the angle between the directions of linear polarizations, and $CP$ is the $CP$ parity of the Higgs boson.

The most reasonable choice of photon collider energy and the laser wavelength for the Higgs study is $E_0=110$ GeV
and $\lambda \sim 1.05$ \MKM\ (most powerful lasers available); the corresponding parameter $x=4E_0 \omega_0/m^2c^4 \approx 2$.

Let us consider the two most important sets of parameters: 1) for the measurement of $\Gamma_{\GG}$, 2) for the measurement of $CP$.
In both cases, it is preferable to use longitudinally polarized electrons, $2\lambda_e=-0.85$ is possible.
For case 1, the laser polarization should be $P_{\mathrm{c}}=1$ and $2P_{\mathrm{c}}\lambda_e \sim -0.85$
(to enhance the number of high-energy photons); then, the resulting  polarization of the scattered photons $\lambda_{1,2} \approx 1$, $l_{1,2}=0$.
For case 2, one should take $P_l=1$, then $\lambda_{1,2}=0.68$, $l_{1,2}=0.6$.
Simulation has been performed for a laser target thickness of 1.35 (in units of the Compton scattering length)
and the CP-IP distance $b=\gamma \sigma_y$; it gave $dL_{0,\GG}/dz/ L_{ee}= 0.84$ and 0.35 for cases 1 and 2, respectively.
The corresponding effective cross sections are 75 fb and 28.5 fb, which should be compared with 290 fb for the process $\EPEM\ \to ZH$.

The geometric $ee$ luminosity in the case of the photon collider is approximately equal to the \EPEM\ luminosity
(the pinch factor in \EPEM\ collisions is compensated by a tighter focusing in \GG\ collisions).
This means that for the same beam parameters the Higgs production rate at the photon collider
is approximately four times lower than in \EPEM\ collisions.

The photon collider can measure better only $\Gamma_{\GG}$, which determines the Higgs
production rate in \GG\ collisions and can be measured by detecting the decay mode
$H\to bb$ ($\sim57\%$ of the total number of Higgs decays).
In \EPEM\ collisions, the Higgs' \GG\
width is measured in the $H \to \GG$ decay, which has a branching fraction of
0.24\%. This means that at the photon collider the statistics for the
measurement of $\Gamma(H\to \GG)$ is higher by a factor of
$0.57/0.0024/4\approx 60$ (or even larger if a lower-emittance
electron source becomes available). This is the main motivation for the
photon collider. The study of the $H\GG$ coupling is arguably the most interesting area of Higgs
physics because it proceeds via a loop and therefore is the most sensitive to New Physics.
The photon collider at the ILC with the expected $L_{ee}\approx 3\times 10^{34}$ will produce
about 22500 Higgs bosons per year ($10^7$ sec), which would enable
the determination of $\Gamma(H\to \GG)\times Br(H\to bb)$ with an accuracy of 2\%
\cite{Krawczyk,Asner,Monig}.

The photon collider can also be used also for the measurement of the Higgs boson's $CP$
properties using linearly polarized high-energy
photons (details are provided below).

As one can see, while \EPEM\ collisions are more powerful overall for the study
of Higgs properties, a \GG\ collider would add very significantly in some areas.
The relative incremental cost of adding a photon collider to an \EPEM\ linear collider
is very low. Therefore, the best solution would be to build an \EPEM\ linear collider
combined with a photon collider; the latter would come almost for free.

  Note, though \epem collisions are certainly better for study of Higgs properties (mainly due to Z-tagging), at high energies the photon collider potential is  close to that in \epem collisions: energies and statistics are similar---but in different reactions~\cite{golden}. In some cases, photon colliders provide access to higher masses or allows the study of some phenomena with higher precision.

\section{The photon collider, optimum energy for the \GG\ Higgs factory}

Photon colliders ($\gg, \gamma e$) based on one-pass linear colliders (PLCs) have been in development since 1981~\cite{GKST81,GKST83}. A detailed description of the PLC can be found in Ref.~\cite{TESLATDR}. After undergoing Compton scattering at a distance $b \sim 1$~mm from the IP (Fig.~\ref{plc}), the photons have an energy close to that of the initial electrons and follow the electrons' original direction toward the IP (with a small additional energy spread of the order of $1/\gamma$). Using a modern laser with a flash energy of several joules, one can ``convert'' almost all electrons to high-energy photons. The maximum energy of the scattered photons (neglecting nonlinear effects) is
\begin{equation}
\omega_m=\frac{x}{x+1}E_0; \;\;\;\;
x \approx \frac{4E_0\omega_0}{m^2c^4}
 \simeq 15.3\left[\frac{E_0}{\tev}\right]\left[\frac{\omega_0}{\ev}\right]=
 19\left[\frac{E_0}{\tev}\right]
\left[\frac{\mum}{\lambda}\right].
\label{x}
\end{equation}

As discussed earlier in the paper, the Higgs boson is produced in \gg collisions as a single resonance.
For the Higgs factory, one can take $E_0\approx 80 \gev$ and $\lambda=1.06/3$ \mum ($x=4.3$) or
$E_0\approx 110 \gev$ and $\lambda=1.06$ \mum ($x \approx 2$). The first option needs a lower electron energy but
is less preferable due to problems with the removal of too-low-energy final electrons that are deflected by
the opposing electron beam and by the detector field.

The luminosity spectrum in \gg collisions has a peak at maximum energies with $\mathrm{FWHM} \sim 15-20\%$ containing the \gg
luminosity of about 10\% of the geometric $e^-e^-$ luminosity. Typical cross sections (Higgs, charged pairs) in \gg collisions are one order
of magnitude higher than in \epem collisions, therefore the numbers of events at photon and \epem colliders are comparable.

Photon colliders are usually considered as an almost free-of-charge addition to a linear \epem collider (now, ILC and CLIC). The ``to be or not to be'' for photon colliders depends on decisions on the underlying LC projects. However, very recently a number of proposals for ``circular'' photon-collider Higgs factories without \epem collisions and without damping rings have been put forward.

 The preferable electron beam energy and laser wavelength for the \GG\ Higgs factory are
 $E_0 \approx 110$ GeV and $\lambda \approx 1\, \MKM$, corresponding to the parameter $x\approx
2$ (this includes the spectrum shift due to nonlinear effects in
Compton scattering). Note that all photon-collider projects that appeared
in the last year assumed $E_0=80$ GeV (85 GeV would be more correct)
and $\lambda=1/3$ \MKM\ ($x=4.6$). This choice was driven by the
simple desire to have the lowest possible collider energy. However, life is
not so simple, there are other important factors that must be considered:

\begin{enumerate}
\item As proposed, these projects would suffer from the very serious problem
of the removal of used electron beams. That is because the minimum energy of electrons
after multiple Compton scattering in the conversion region will be
a factor of 4.5 lower~\cite{TESLATDR}, and these electrons will be deflected
at unacceptably large angles by the opposing beam as
well as by the solenoid field (the latter due to the use of the crab-crossing collision scheme).

\item For the measurement of the Higgs' $CP$ properties one should collide
linearly polarized $\gamma$ beams at various angles between
their polarization planes. The effect is proportional to
the product of linear polarizations $l_1l_2$. The degree of linear
polarization at the maximum energies is 60\% for $x=2$ and 34.5\% at
$x=4.6$. This means that the effect in the latter case will be 3
times smaller, and so in order to get the same accuracy one would have to
run the experiment 9 times longer.
The case of $x=1.9$ was simulated, with backgrounds taken into account, in ref.~\cite{Asner};
it was found that the $CP$ parameter (a value between 1 and $-1$) can be measured with a 10\% accuracy
given an integrated geometric $ee$ luminosity of $3\cdot 10^{34} \times
10^7$ = 300 fb$^{-1}$.
\end{enumerate}
Both of these facts strongly favor a photon collider with $E_0=110$ GeV and $\lambda \approx 1$ \MKM.

\section{Photon colliders at the ILC and CLIC}

The future of these collider projects is quite unclear due to their
high cost, complexity, and (as of yet) absence of new physics in
their energy region (other than the Higgs boson). If ILC in Japan
is approved, there is a very high probability that it will
include the photon collider.

The photon collider for TESLA (on which ILC is based) was considered in detail at the conceptual
level~\cite{TESLATDR,telnov}. The next major step must be R\&D for
its laser system, see Sect.\ref{lasers}.

The expected \EPEM\ luminosity of the updated ILC design at $2E_0=250$ \gev is
$3\cdot 10^{34}\ \cms$. The geometric $ee$ luminosity at the \GG\ collider
could be similar. To further increase the \GG\ luminosity, one needs
new ideas on the production of low-emittance polarized electron beams.
ILC damping rings are already close to their ultimate performance.
To increase the luminosity further, I have proposed~\cite{telnov2} to combine
many (about 50-100) low-charge, low-emittance
bunches from an RF photogun into a single bunch in the longitudinal phase
space using a small differential in beam energies.
Using this approach, it may be possible to
increase the luminosity by a factor of 10 compared to that with damping
rings. To achieve this, we need low-emittance polarized RF guns, which have
appeared only recently and are yet to reach their ultimate
performance. In the past, only DC polarized photoguns were available, which
produce beams that require further cooling with damping rings. The idea of
beam combining is highly promising and needs a more careful consideration.

The TESLA TDR, published in 2001, dedicated a 98-page chapter to
the photon collider. The recently published ILC TDR, on the other hand,
includes only a brief mention of the photon collider, as an option.
The scope document on linear colliders, developed and supported by the physics
community, states that the ILC design should be compatible with the
photon collider. The focus of the present ILC TDR was the minimization of cost
while attempting to preserve ILC's primary performance characteristics.
This has resulted in cuts in all places possible. In particular,
only one IP remains in the design, instead of two, with two pull-push detectors.
In the ILC TDR, the IP was designed for a beam crossing angle of 14 mrad,
while the photon collider requires a crossing angle of 25 mrad.
The choice of a crossing angle incompatible with the photon collider
was made simply because all attention in the TDR effort was focused on the
baseline \EPEM\ collider, not because someone was against the photon collider
(no one was). It is not too late to reoptimize the ILC IP and make it compatible
with the photon collider.

\section{Photon colliders based on recirculating linacs}
\begin{figure}[!htb]
\centering
\includegraphics[width=10.cm,height=6.5cm,angle=0]{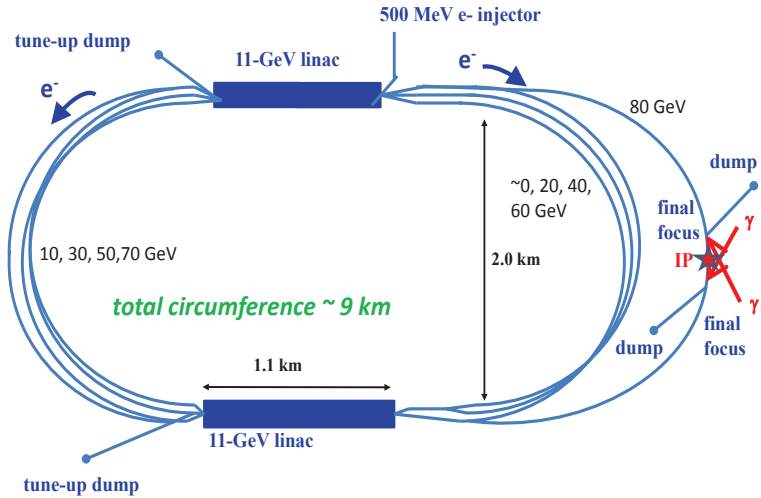}
\caption{The SAPPHiRE Higgs factory } \label{sapphire}
\end{figure}

  About one year ago, F.~Zimmermann et al.~\cite{Sapphire} proposed to use
the 60 GeV recirculating electron linac developed for $ep$ collisions with
LHC protons (LHeC) as a photon collider (project SAPPHiRE).
The ring contains two 11 GeV superconducting linacs and six arcs, each
designed for its own beam energy.
An injected electron would make three turns
to reach the energy of 60 GeV required for LHeC. To obtain the 80 GeV required
for the photon collider, the authors propose adding two additional
arcs, see Fig.~\ref{sapphire}. One must also double the number of arcs
to accommodate the
second electron beam traveling in the opposite direction. It was proposed
to use polarized electron beams with no damping rings; the required photoguns
are still under development.

  In any case, the idea is interesting because two 80 GeV electron beams
are obtained with only 22 GeV's worth of linac.
The radius of arcs is 1 km, and the total circumference is 9 km.
On the other hand, the total length of all arcs is 72 km!
In addition, such a design would be impractical due to the unacceptable increase of horizontal
emittance in the bending arcs (about degradation of the vertical emittance see below). The increase of the normalized emittance per turn is proportional to $E^6/R^4$.
To solve this problem, the authors of SAPPHiRE have proposed to use $\times 4$
shorter arc structures, which would lead to $\times 64$ smaller emittance
dilution. This might be possible but would require $\times 16$
stronger quadrupole magnets.

  Another weak point of this proposal is the use of 80 GeV electron beams and the 1/3
  \MKM\ laser wavelength. As mentioned above,
  this choice of parameters makes it very difficult to remove the
  disrupted electron beams from the detector
 and leads to low sensitivity in the measurement of the
  $CP$ properties of the Higgs boson.

  It is highly unlikely that the LHeC project (and, correspondingly, SAPPHiRE) will be approved.
  However, the idea behind SAPPHiRE has become very popular and has been cloned for all existing
  tunnels at major HEP laboratories. In particular, it has been proposed to
  build a photon collider in the Tevatron ring at FNAL (6 km circumference),
Higgs Factory in Tevatron Tunnel (HFiTT) ~\cite{FNAL}.
  This collider would contain 8 linac sections providing a total energy gain of
  10 GeV per turn. In order to reach the energy of 80 GeV,
  the electron beams would make 8 turns. The total number of beamlines in
  the tunnel will be 16, with the total length of approximately 96 km. This
  proposal contains just a desired set of numbers without any attempt at
  justification. Simple  estimates show that such a collider will not work due to the strong
  emittance dilution both in the horizontal and vertical directions. The
  eight arcs would be stacked one on top another, so electrons will jump up and down, by up to 1.5 m, 16
  times per turn, 128 times in total. The vertical emittance is
  assumed to be same as in the ILC damping ring; it will be certainly destroyed on such ``mountains''.

\begin{figure}[!htb]
\centering
\includegraphics[width=10.cm,height=6.5cm,angle=0]{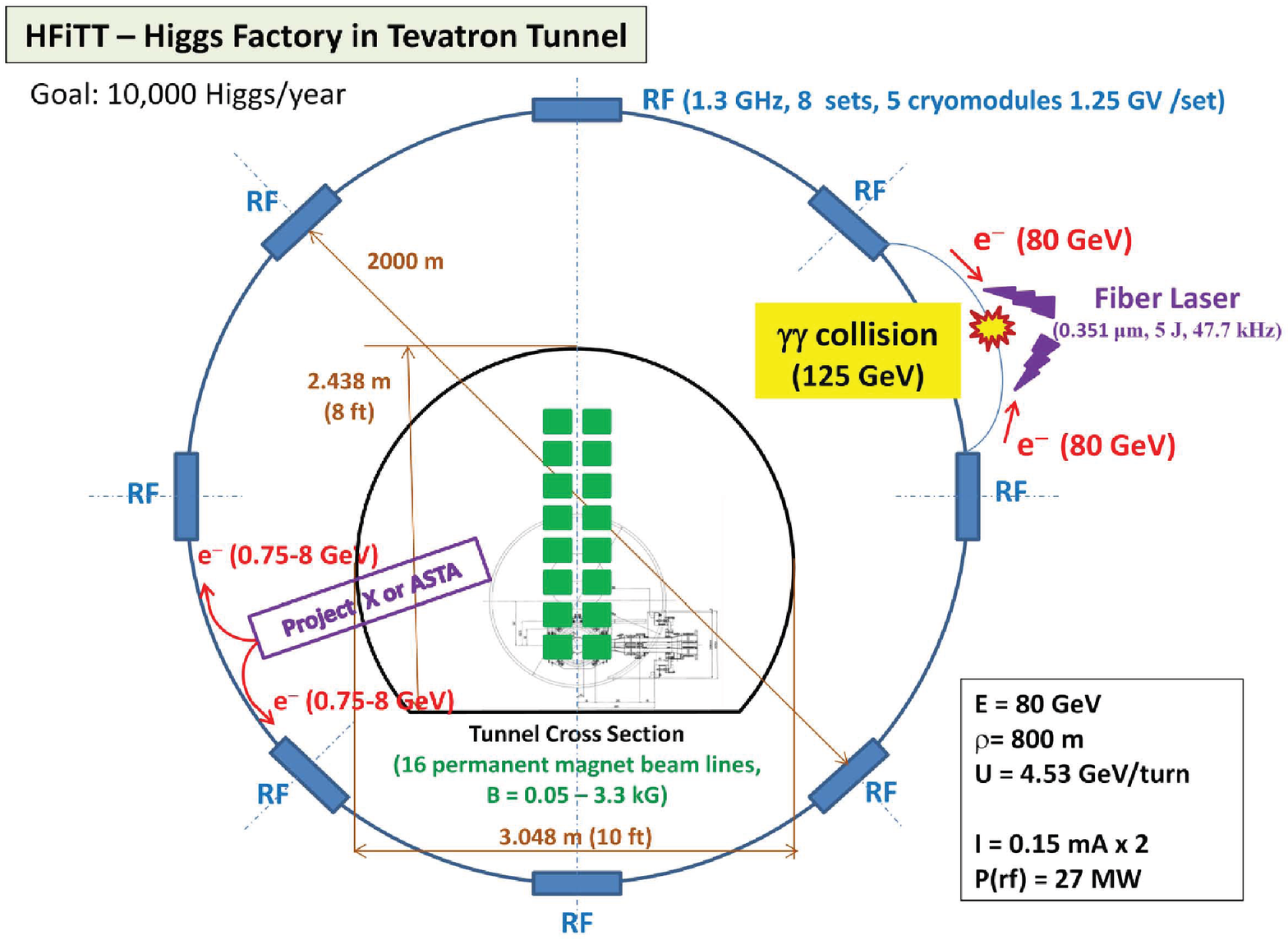}
\vspace*{0.5cm}
\caption{The HFiTT Higgs factory} \label{fnal}
\vspace*{0.5cm}
\end{figure}

 The most interesting feature of the HFiTT proposal is a novel laser
  system based on fiber lasers, see Sect.~\ref{lasers}.

\begin{figure}[tbph]
     \begin{center}
     \vspace*{0.5cm}
\includegraphics[width=12cm] {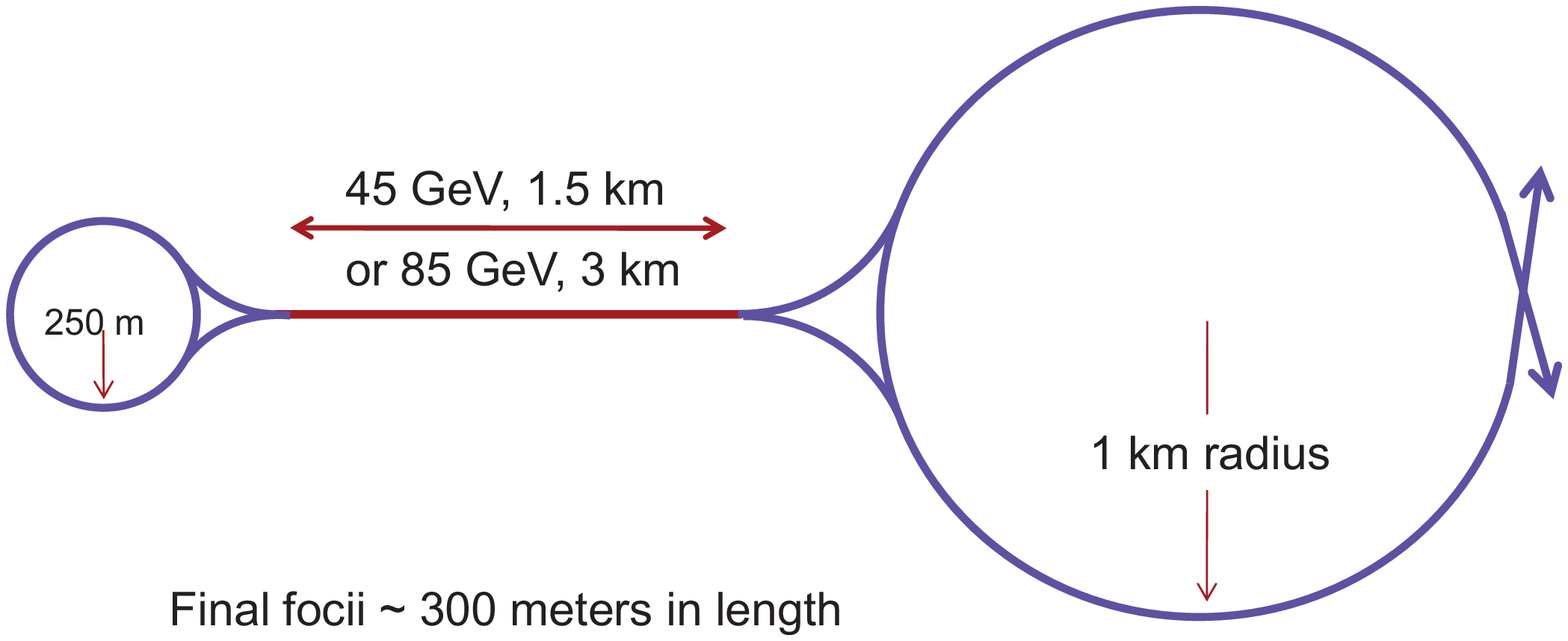}
       \vspace*{0.cm}
     \end{center}
     \caption{The SLC-type photon collider.}
   \vspace*{0.0cm}
   \label{SLC}
   \end{figure}

  There is also a proposal~\cite{raubenheimer} to build a photon collider based on the existing
  SLAC linac. Electrons would acquire 40 GeV traveling in the linac in
  one direction, then make one round turn in a small ring, get another
  40  GeV traveling in the same linac in the opposing direction, and then the two
  beams would collide in $R=1$ km arcs, similar to the SLC, Fig.~\ref{SLC}..

  It is a nice proposal; however, for the Higgs factory it is desirable
  to have $E_0=110$ GeV, as explained above. Reaching 110 GeV would require
  either a higher acceleration gradient (or an additional
  30 GeV injector) and arcs with a larger radius.

 \section{Laser systems for photon colliders} \label{lasers}

\begin{figure}[!tbph]
     \begin{center}
\includegraphics[width=13cm] {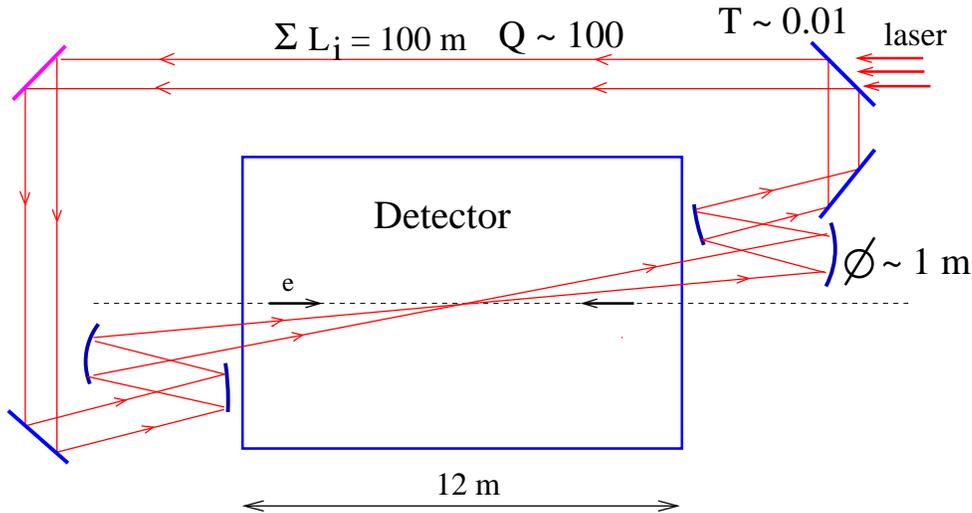}
       \vspace*{0.0cm}
     \end{center}
     \caption{Optical cavity laser system for the photon collider at ILC.}
   \vspace*{0.cm}
   \label{cavity}
   \end{figure}

The requirements on the laser system for the PLC are as follows: flash energy about 10 J, duration $\sim$ 1 ps, the wavelength $\lambda \sim 1$ \mum, and the pulse structure similar to that for the electron beams. In the case of single use of laser pulses, the average energy of each of the two lasers should be of the order of 100 kW, both for ILC and CLIC. At the ILC, the distance between the bunches is large, about 100-150 m, which makes possible the use of an external optical cavity (Fig.~\ref{cavity}), which, in turn, can reduce the required laser power by a factor of $Q \sim 100$. At CLIC, the distance between the bunches is only 15 cm, and so having an optical cavity is not possible---one needs a very powerful one-pass laser.

Such a laser system, while certainly feasible, would not be easy
to build and would require a great deal of R\&D and prototyping. The optical-cavity technology, proposed for the photon collider in 1999,
has been developed very actively for many applications based on Compton
scattering; however, its present status is still far from what is needed
for the photon collider.

\begin{figure}[htp]
     \begin{center}
     \vspace*{-0.2cm}
\includegraphics[width=7.3cm] {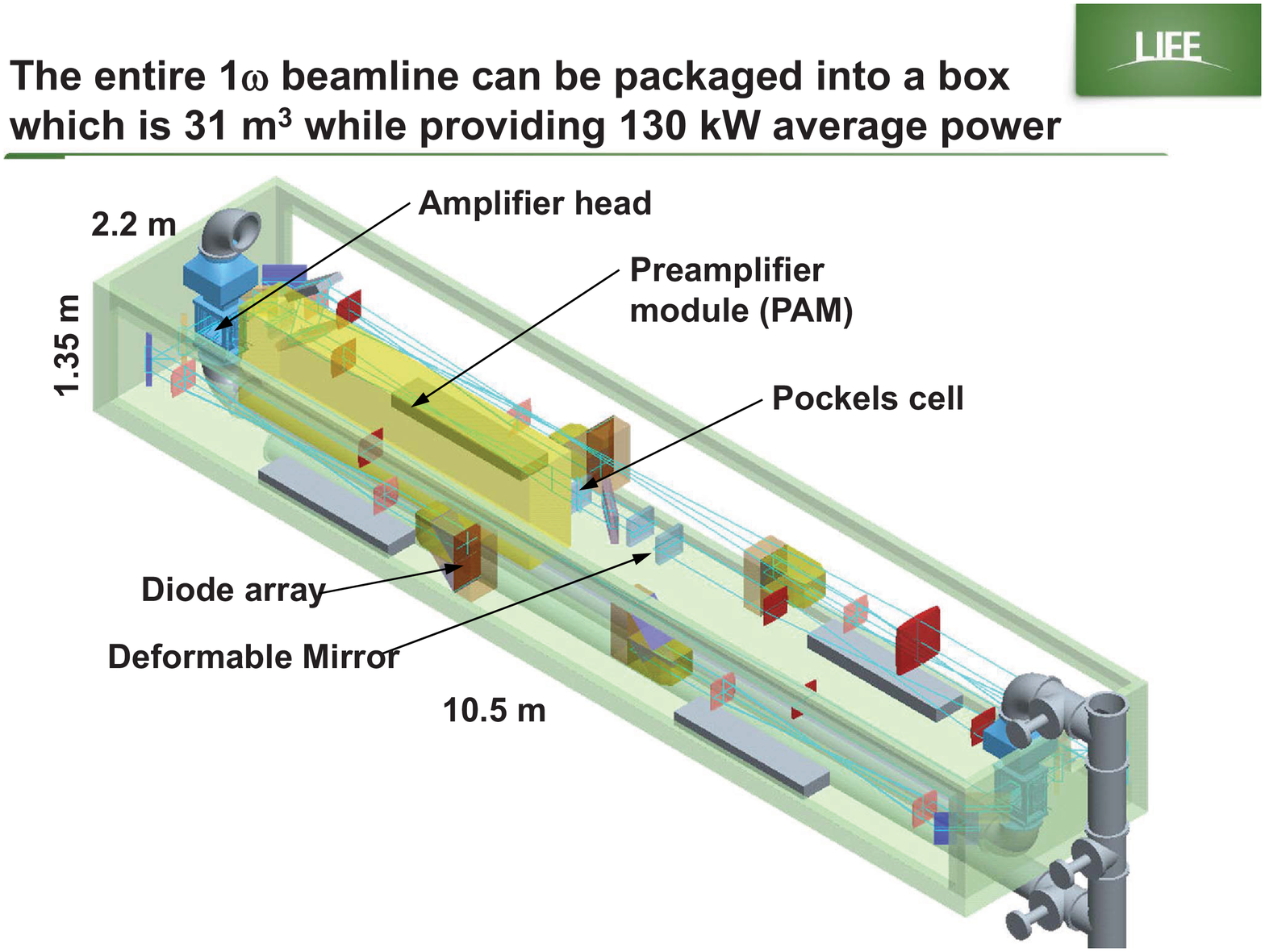} ~~ \includegraphics[width=7.3cm] {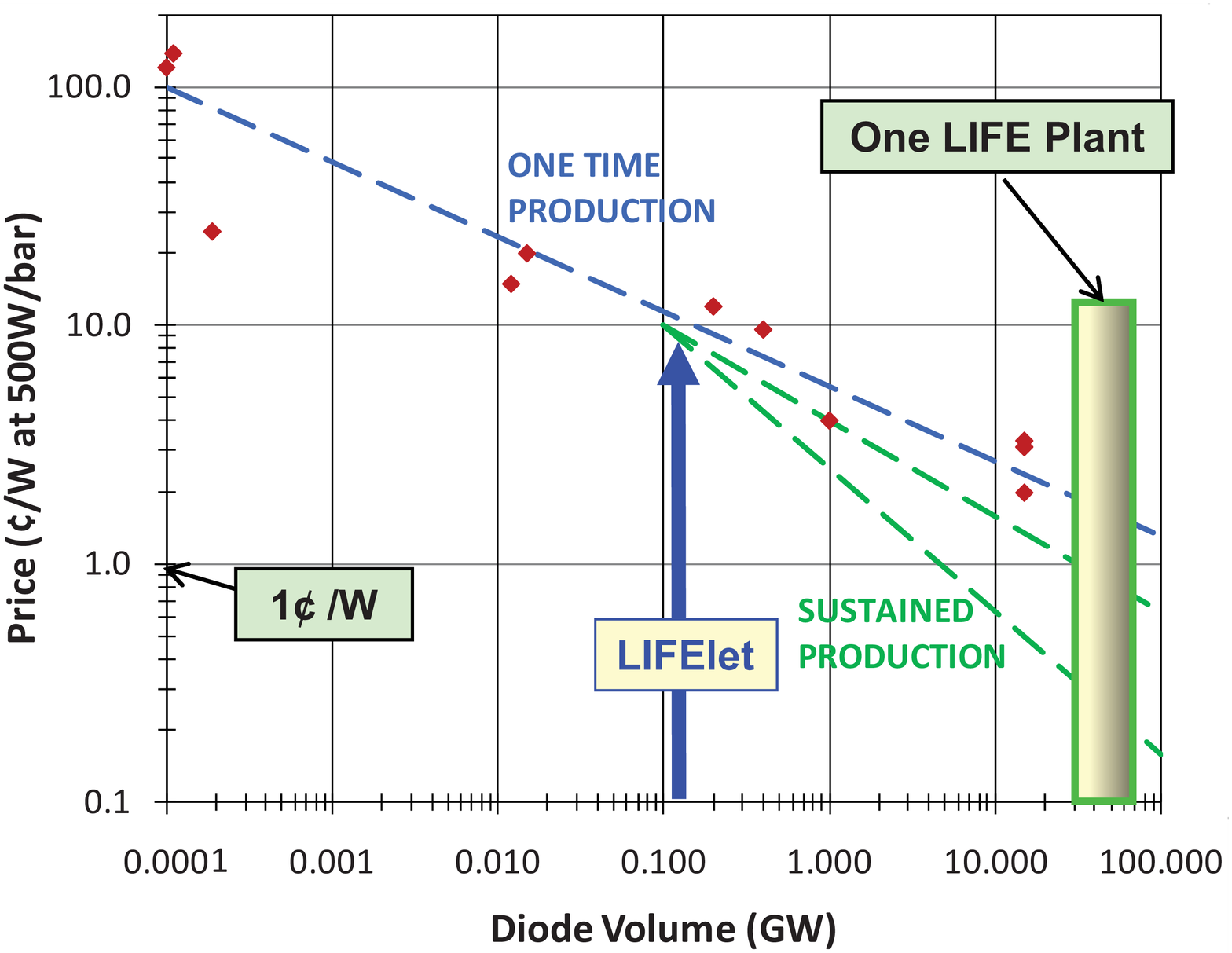}
       \vspace*{-0.cm}
     \end{center}
     \caption{Left: One laser for the inertial fusion project LIFE at LLNL, suitable for photon colliders at both ILC and CLIC. Right: industrial price of diodes per one watt.}
   \vspace*{0.5cm}
   \label{LLNL}
   \end{figure}

 New hopes came from LLNL's laser-fusion project LIFE~\cite{Bayramian}, which is based
 on the diode-pumping technology. LIFE's laser system will consist of about
 300 lasers, each operating at a repetition rate of 16~Hz and delivering
 8.4 kJ per flash. The photon collider at the ILC would require a laser
 that produces 1 ms trains of 2600 pulses, 5-10 J per pulse, with a
 repetition rate of 5-10 Hz.  LLNL experts say that the LIFE laser can
 be modified for the production of the required pulse
 trains with further chirped pulse compression. Fig.~\ref{LLNL} (left) shows one of LIFE lasers. Its volume is $31 \mathrm{m}^3$. The advancement of this
 technique has been enabled by the significant reduction of the cost of
 pumping diodes, currently estimated at \$0.10 per watt,
 which translates to \$3 million per laser (the ILC-based photon collider
 would require $\sim 6$ such lasers). If so, one can use such a laser both for ILC and CLIC without any enhancement in optical cavities.

 Naturally, it is very  attractive to simply buy a few \$3M lasers and use them in one-pass mode
 rather then venturing to construct a 100 m optical cavity and stabilize
 its geometry with an accuracy of several nanometers. For the CLIC-based
 photon collider, the optical-cavity approach would not work at all due to
 CLIC's very short trains; a LIFE-type laser is therefore the only viable option.

\begin{figure}[htp]
     \begin{center}
     \vspace*{-0.cm}
\includegraphics[width=12cm] {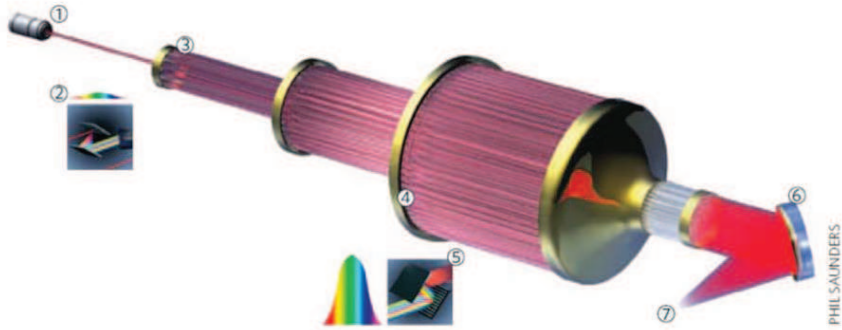}
       \vspace*{-0.3cm}
     \end{center}
     \caption{Fiber laser system ~\cite{FNAL}.}
   \vspace*{0.0cm}
   \label{FNAL-l}
   \end{figure}

 Another very promising laser system for photon colliders was suggested for the HFiTT photon collider project discussed above.
Only recently have laser physicists  succeeded in coherently combining the light from thousands of fibers.
  A diode-pumped fiber laser, Fig.~\ref{FNAL-l}, is capable of producing 5-10 J pulses with a repetition rate of 47.7 kHz
  as required by HFiTT.  In this project electron bunches follow with constant repetition rate that is very convenient for such laser system because pumping diodes work continuously that reduce their total pulse power.

  It would have been very attractive to use such a fiber laser for
  the photon collider at the ILC as its total power would be larger than
  needed. Unfortunately, the pulse structure at the ILC would be very bad for a such
  laser, as the ILC needs $2600\times 10$ J $=26$ kJ per 1 ms, which translates to a 55
  times greater (peak) power of the diode system. Correspondingly,
  the diode cost would be greater by the same factor.
\section{Conclusion}

  The photon collider based on ILC (or CLIC) is a most realistic
  project. However, if the \EPEM\ program occupies all the experiment's time,
  the photon collider will not become reality for least 40 years from now, which is
  unattractive for the present generation of physicists. More attractive would be
  a collider with two interaction regions.

  A laser system based on the project LIFE lasers is the most attractive choice at this time;
  fiber lasers can also reach the desired parameters at some point in future.
  Development of low-emittance polarized electron beams can
  increase the photon collider luminosity by a further order of
  magnitude. The photon collider would be very useful for the precise measurement
  of the Higgs' \GG\ partial width and its $CP$ properties. A very high-luminosity photon
  collider at the energy $2E_0=400$ GeV can help measure the Higgs' self coupling. The
  photon collider based on ILC (CLIC) can work with the 1 \MKM\ laser
  wavelength up to $2E_0 \sim 700$ GeV; for higher energies, one should use a greater laser wavelength.

  The idea of a photon-collider Higgs factory based on recirculating linacs
  looks interesting as it can use shorter linacs. Unfortunately, the problem of
  emittance dilution is very serious and the total length of the arcs is very large.
  The pulse structure of such colliders (equal distance between collisions) is very well suited for fiber
  lasers. Such a recirculating collider with a desirable $E_0$ $(\approx 110$ GeV) can possibly work in large rings such as
  LEP/LHC or UNK, but then the total length of arcs will be several hundred km and the cost would exceed
  that for linear colliders with similar energy (that could be, for example, a warm linear collider with the 4 km length).
  Most importantly, a photon collider with no \EPEM\ does not make much sense for the study of the Higgs boson.
  At this time, the ILC is the best place for the photon collider. Unfortunately, future of linear colliders remain uncertain already several decades.

\section*{Acknowledgments}

The work was supported by the Ministry of Education and Science of the Russian Federation.

\end{document}